\documentclass[iop]{emulateapj}

\newcommand{\kev}{keV}
\newcommand{\fe}{Fe~K$\alpha$}
\newcommand{\etal}{et al.}
\newcommand{\lmc}{LMC~X-4}
\newcommand{\oeight}{O~\textsc{viii}}
\newcommand{\suzaku}{\textit{Suzaku}}

\newcommand{\xmm}{\textit{XMM-Newton}}

\shorttitle{X-ray Reflection \& X-ray Pulsars}
\shortauthors{Ballantyne \etal}

\usepackage{times}

\begin{document} 

\title{Ionized Reflection Spectra from Accretion Disks Illuminated by X-ray Pulsars}


\author{D. R. Ballantyne\altaffilmark{1},
  J. D. Purvis\altaffilmark{1}, R. G. Strausbaugh\altaffilmark{1}, and
R. C. Hickox\altaffilmark{2}}
\altaffiltext{1}{Center for Relativistic Astrophysics, School of Physics,
  Georgia Institute of Technology, Atlanta, GA 30332; david.ballantyne@physics.gatech.edu}
\altaffiltext{2}{Department of Physics and Astronomy, Dartmouth
  College, 6127 Wilder Laboratory, Hanover, NH 03755}

\begin{abstract}
X-ray reflection signatures are observed around multiple classes of
accreting compact objects. Modelling these features yield
important constraints on the physics of accretion disks, motivating
the development of X-ray reflection models appropriate for a variety
of systems and illumination conditions. Here, constant density ionized X-ray reflection
models are presented for a disk irradiated with a very hard power-law
X-ray spectrum ($\Gamma < 1$) and a variable high-energy cutoff. These
models are then applied to the \suzaku\ data of the accreting X-ray
pulsar \lmc, where very good fits are obtained with a highly ionized
reflector responsible for both the broad \fe\ line and the soft
excess. The ionized reflector shows strong evidence for significant
Doppler broadening and is redshifted by $\sim 10^4$~km~s$^{-1}$. These
features indicate that the reflecting material is associated with the
complex dynamics occurring at the inner region of the
magnetically-truncated accretion disk. Thus, reflection studies
of X-ray pulsar spectra may give important insights into the accretion
physics at the magnetospheric radius.
\end{abstract}


\section{Introduction}
\label{sect:intro}
Many astrophysical systems contain a hard X-ray source in close
proximity to dense, relatively cold gas. In many cases the X-rays
illuminate the dense gas, which scatters and absorbs the energy before
ultimately emitting a spectrum that is imprinted with a variety of diagnostic
spectral features, most notably the \fe\ fluorescence line \citep[e.g.,][]{lw88,gr88,gf91,mpp91}. Such an X-ray
reflection spectrum has been observed from accretion disks around
numerous neutron stars \citep[e.g.,][]{cack10} and black holes
of all sizes \citep[e.g.,][]{mill07}. In these systems, modeling the observed X-ray reflection features
can lead to important constraints on the ionization stage
\citep{rfy99,rf05,gk10}, metallicity \citep{ball02}, and density structure \citep{bal01,ball04,ball05} of the inner accretion
disk. In addition, the reflecting region of the disk may be subject to
relativistic effects that will sculpt the emission features \citep[e.g.,][]{fab89,lao91}
leading to measurements of disk radii and black hole spin \citep{br06,mill07,rf08}. It is clear that modeling X-ray reflection features is a very
important method in understanding the physics of accreting compact
objects; however, such fundamental measurements are only possible by comparing
the observed spectra to grids of models that are computed for the
appropriate radiative conditions of the system under consideration.

Here, we calculate ionized X-ray reflection models appropriate for
accretion powered X-ray pulsars. X-ray pulsars are a class of X-ray
binaries in which the neutron star has a strong magnetic field, so
that accreting matter follows the field lines and falls into the
magnetic poles, generating pulses as the magnetic poles rotate in and
out of our line-of-sight \citep[e.g.,][]{naga89,bild97}. The inner
accretion disk is generally believed to be truncated at the
magnetospheric or Alfv\'{e}n radius, where the energy density of the
system is dominated by the magnetic field (e.g.,
\citealt{fkr02}). Reflection of hard X-rays from the neutron star by
the inner disk has been widely invoked to explain the ubiquitous soft
excess component observed in X-ray pulsars (see \citealt{hnk04} and
references therein) as well as fluorescent iron emission lines at
$\approx 6.4$--6.7~keV
\citep[e.g.,][]{choi94,woo95,dalf98,burd00,lab01,endo02,naik03,naik04}.
Despite the ubiquity of these apparent reflection features in
observations of X-ray pulsars, there has been little work aimed at
fitting the spectra with a self-consistent, physically-motivated
reflection model, which is the goal of the present study.  The next
section briefly describes the calculation of the grid of reflection
models that can be used for X-ray pulsar data, which is then used in
Section~\ref{sect:res} to fit the \suzaku\ data of \lmc\
\citep{hung10}. A summary of our conclusions is found in
Section~\ref{sect:concl}.

\section{Calculations}
\label{sect:calc}
Descriptions of the reflection calculations have been previously
published \citep[e.g.,][]{rf93,rfy99,rf05}, so only the main features of the model will be
presented here. A one-dimensional, constant density slab of gas with
hydrogen number density $n_{\mathrm{H}}$ is illuminated by a X-ray
continuum with flux $F_{X}$ defined between $1$~eV and
$100$~keV. Thus, for a given spectral shape, models can be
described by an ionization parameter
\begin{equation}
\xi = {4 \pi F_{\mathrm{X}} \over n_{\mathrm{H}}}.
\label{eq:xi}
\end{equation}
Hydrogen and helium are assumed to be fully ionized (i.e.,
$n_{e}=1.2n_{\mathrm{H}}$), and the treated metals (C, N, O, Mg, Si, and
Fe) have the abundances of \citet{mcm83}. The slab has a constant Thomson
depth of $\tau_{\mathrm{T}}=15$ ensuring that all X-ray photons
interact several times even at the largest values of $\xi$. The
calculation finishes, and outputs the angle-averaged reflection spectrum, when all 110 zones of the slab reach thermal and ionization
balance. The ionization stages of the metals included in the
calculation are C~\textsc{v}--\textsc{vii}, N~\textsc{vi}--\textsc{viii},
O~\textsc{v}--\textsc{ix}, Mg~\textsc{ix}--\textsc{xiii},
Si~\textsc{xi}--\textsc{xv} and Fe~\textsc{xvi}--\textsc{xxvii}.

The continua of X-ray pulsars can be described as a
hard power-law with a high-energy cutoff:
\begin{equation}
f_E \propto \left \{ \begin{array}{ll}
 E^{-\Gamma} & \mbox{$E < E_{\mathrm{cut}}$} \\
 E^{-\Gamma} e^{(E_{\mathrm{cut}}-E)/E_{\mathrm{fold}}} & \mbox{$E \geq
   E_{\mathrm{cut}}$},
\end{array}
\right.
\label{eq:highecut}
\end{equation}
where $f_E$ is the photon flux per energy interval, $\Gamma$ is the
photon index, $E_{\mathrm{cut}}$ is the cutoff energy, and
$E_{\mathrm{fold}}$ is the e-folding energy. The cutoff energies for
most sources are typically $< 30$~\kev\ \citep{hnk04}, much less than
the $>100$~\kev\ cutoffs assumed in the publicly available power-law
reflection models \citep[e.g.,][]{bal01,rf05}. Similarly, the photon
indices of X-ray pulsars are $\Gamma \la 1$, significantly harder
than the reflection models calculated for fitting AGNs \citep{rf05}. These
very hard continua also indicate that the reflection spectra will be
insensitive to the 1~eV lower limit assumed in the calculation.

\begin{figure}
\includegraphics[angle=-90,width=0.5\textwidth]{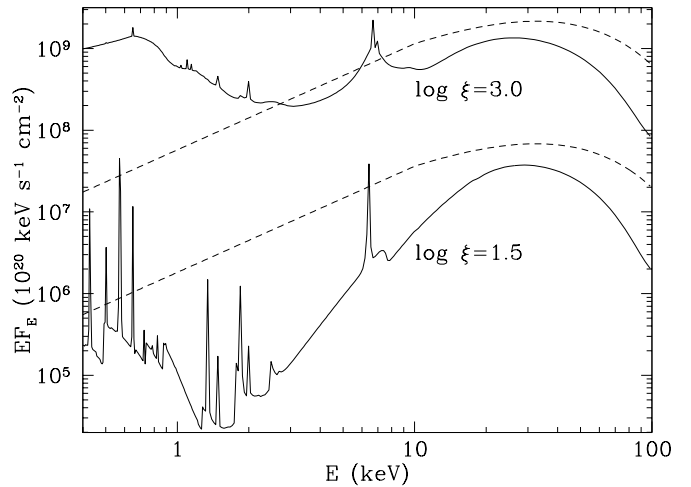}
\caption{Reflection spectra from a constant density slab irradiated by
a power-law with a high-energy cutoff (Eq.~\ref{eq:highecut}) with
$\Gamma=0.7$, $E_{\mathrm{cut}}=10$~keV and
$E_{\mathrm{fold}}=25$~keV (dashed lines). The solid lines plot the
reflection spectra for both a weakly ionized slab ($\log \xi=1.5$) and
a strongly ionized one ($\log \xi=3.0$). The highly irradiated slab
produces a strong bremsstrahlung dominated soft excess at energies $\la 3$~keV.}
\label{fig:model}
\end{figure}

A grid of 2376 models is calculated with parameters $5$~keV$ \leq
E_{\mathrm{cut}} \leq 30$~keV (in steps of $5$~keV), $5$~keV$ \leq
E_{\mathrm{fold}} \leq 30$~keV (in steps of $5$~keV), $0.5 \leq \Gamma
\leq 1.5$ (in steps of $0.1$) and $1.5 \leq \log \xi \leq 4.0$ (in
steps of $0.5$). The ionization parameter is varied by fixing the
density at 
$n_{\mathrm{H}}=10^{19}$~cm$^{-3}$ (appropriate for $r \sim 20$~$r_g$
in a radiation pressure dominated accretion disk around a neutron
star; $r_g=GM/c^2$, where $M$ is the mass of a neutron star) and
changing $F_X$\footnote{The exact value of the density at the surface of a
  X-ray pulsar accretion disk is unknown (and will likely be
  functions of the disk radius and accretion rate; e.g.,
  \citealt{ss73}), but is expected to be $n_{\mathrm{H}} \gg
  10^{20}$~cm$^{-3}$, a region where calculations become
  computationally prohibitive. Recall, however, that the
$\alpha$-disk model gives the midplane or an average density, and the
density at the surface is likely to be much lower.}. Two reflection spectra with different $\xi$ are shown
in Figure~\ref{fig:model}. The features in this spectra are qualitatively very similar to those used in modeling
observations of AGNs. However, because of the high density in
neutron star disks and the very hard illuminating power-law, the $\log
\xi=3$ model predicts a strong bremsstrahlung dominated soft excess at
energies $\la 3$~keV from the hot surface of the X-ray heated
disk. The thermal bremsstrahlung continuum is
modified by Comptonization in the hot atmosphere so that it
approximately resembles the blackbody spectrum typically used to fit
the observed soft excess. In addition, the low ionization model predicts several strong
recombination lines from ionized C, N and O in the slab which, if
measured, could constrain the abundances of the accreting
gas. Interestingly, the soft excesses in X-ray pulsar spectra are
observed to be featureless in observations with CCD resolution \citep{hnk04}; therefore, if X-ray
reflection is responsible for the soft excess, the ionization
parameter of the reflector must be large.

\section{Application to LMC X-4}
\label{sect:res}
To determine if the ionized reflection models provide a good
description of the soft excesses observed from X-ray pulsars, the grid
of models described above is fit to three phase-averaged \suzaku\
spectra of \lmc\ \citep{hung10}. LMC X-4 is a high-mass X-ray binary
system consisting of a 1.29 $M_{\sun}$ neutron star accreting from a
14.5 $M_{\sun}$ O8 III companion \citep{kell83,meer07,rawls11}.  The neutron
star rotates with a period of $\sim$13.5~s, orbits its companion with
a period of $\sim$1.4~days \citep{whit78}, and exhibits a long-term
periodicity of $30.5\pm0.5$ days that is believed to be
caused by a precessing accretion disk that periodically obscures the
neutron star \citep[e.g.,][]{lang81,heem89,naik03}.

LMC X-4 was observed by the \suzaku\ X-ray observatory in 2008
January--April, with exposures of $\approx$20 ks duration taken at
three different phases in the superorbital cycle ($\phi_{30} = 0.39$,
0.27, and 0.07 for Observations 1, 2, and 3, respectively). Details of
the data reduction and spectral extraction are presented in
\citet{hung10}. Phase-averaged spectra were extracted in the 0.5--10
keV energy range for the X-ray Imaging Spectrometer (XIS) CCD
detector, and the 10--50 keV range for the PIN instrument of the Hard
X-ray Detector (HXD), and the spectra fitted jointly for each
observation. \citet{hung10} found that the broad-band \suzaku\ spectra
are well described by a hard power-law with a high-energy cutoff, a
soft X-ray excess, and a strong iron emission line, similar to
previous studies \citep[e.g.,][]{woo96,lab01,naik03}. The best-fit
models also included emission features at $\approx 0.6$~keV and
$\approx 1$~keV that were identified by \citet{hung10} as emission from highly
ionized O and Ne, respectively. An analysis of the energy-resolved pulse
profiles showed that the phase offset between hard ($>2 $~keV) and
soft ($< 1$~keV) pulsations varies with time, as has been observed for
the similar X-ray pulsars Her X-1 \citep{rams02, zane04} and SMC X-1
\citep{neil04, hick05}. As discussed in \citet{hung10}, this behavior
is consistent with the origin of the soft component as the illuminated
inner region of the precessing accretion disk.

In this paper we further explore the physical origin of the reflected
emission in LMC X-4, by re-analyzing the {\em Suzaku} spectra using
our reflection model.  XSPEC v.12.7 \citep{arn96} is used to fit the
\suzaku\ data. Errorbars on the best fit parameters are the 2$\sigma$
uncertainties for one parameter of interest (i.e., $\Delta
\chi^2=2.71$).

\subsection{Spectral Fits}
\begin{figure}
\includegraphics[angle=-90,width=0.5\textwidth]{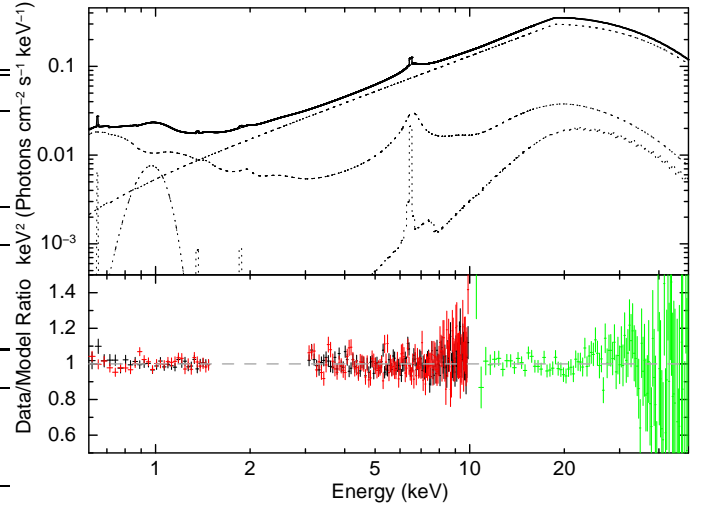}
\caption{The best-fitting model (top panel) and data-to-model ratio
  (lower panel) found for Observation 1 of \lmc. In the upper panel, the thick solid line
  plots the total model, while the individual components (the
  continuum, the reflectors and the Gaussian emission line) are
  indicated with dotted lines. In the bottom panel, the black/red/green data
  points are from the \suzaku/FI/XIS1/PIN detectors, respectively. }
\label{fig:fit3}
\end{figure}

The spectral model used in the fitting is the same as the one used by
\citet{hung10} except the blackbody and emission lines are
replaced by the ionized reflection model. \citet{hung10} found that
both a narrow and broad \fe\ line were required to fit the
\suzaku\ data. Therefore, two reflectors are used in the new spectral
model, with one subject to Doppler broadening using `gsmooth'. The two reflection components and the primary power-law
all have the same $\Gamma$, $E_{\mathrm{cut}}$ and
$E_{\mathrm{fold}}$. The ionization parameter of the unbroadened
reflection spectrum is fixed at $\log \xi=1.5$ so that it can account
for the observed narrow, neutral \fe\ line, but its normalization is
allowed to vary. The results of all three
fits are shown in Table~\ref{table:results} and show that in every case the ionized
reflection model is an excellent description of the broadband spectrum
of \lmc\ (see Figure~\ref{fig:fit3}). In order to fit the centroid of
the broad \fe\ line at $\sim 6.5$~\kev\ \citep{hung10}, the
ionized reflection spectrum (which produces a $6.7$~\kev\ line from
He-like iron) must be redshifted by $\sim
10^4$~km~s$^{-1}$. Interestingly, the fits require an additional Gaussian line with an equivalent width (EW) of $\sim
100$~eV at an energy of $\approx 0.94$~keV. The possible origin of
this line is discussed below.
\begin{deluxetable}{cccc}
\tablewidth{3.5in}
\tablecaption{\label{table:results}\lmc\ Spectral Fitting Results}
\tablecolumns{4}
\tablehead{
\colhead{Parameter} & \colhead{Observation 1} & \colhead{Observation 2} &
\colhead{Observation 3}} 
\startdata
$\chi^2_{\nu}$ (385 d.o.f.) & 1.13 & 1.22 & 1.04 \\
$N_{\mathrm{H}}$ ($\times 10^{22}$~cm$^{-2}$) (fixed) & 0.057 & 0.057
& 0.057 \\
$F_{\mathrm{0.6-50\ keV}}$ (erg cm$^{-2}$ s$^{-1}$) & $8.6\times
10^{-10}$ & $1.4\times 10^{-9}$ & $1.1\times 10^{-9}$ \\
\cutinhead{Continuum}
$\Gamma$ & $0.68^{+0.01}_{-0.02}$ & $0.67\pm 0.01$ & $0.67\pm 0.02$ \\
$E_{\mathrm{cut}}$ (keV) & $18.7\pm 0.5$ & $18.0\pm 0.4$ &
$17.9^{+0.6}_{-0.7}$ \\
$E_{\mathrm{fold}}$ (keV) & $13.3\pm 0.5$ & $15.6^{+0.5}_{-0.4}$ &
$14.9\pm 0.7$ \\
\cutinhead{Redshifted \& Doppler Broadened Ionized Reflection Spectrum}
$\log \xi$ & $3.02\pm 0.01$ & $2.99\pm 0.01$ & $3.02\pm 0.01$ \\
$z$ & $0.034^{+0.011}_{-0.003}$ & $0.034^{+0.004}_{-0.009}$ & $0.036^{+0.012}_{-0.021}$ \\
$\sigma_{\mathrm{6\ keV}}$ (keV) & $0.19^{+0.07}_{-0.05}$ & $0.18^{+0.06}_{-0.05}$ & $0.37^{+0.16}_{-0.21}$ \\
\cutinhead{Unblurred Reflection Spectrum}
$\log \xi$ (fixed) & $1.5$ & $1.5$ & $1.5$ \\
\cutinhead{Gaussian Emission Component}
$E$ (keV) & $0.94\pm 0.01$ & $0.95\pm 0.01$ & $0.94\pm 0.01$ \\
$\sigma$ (keV) & $0.11\pm 0.01$ & $0.10\pm 0.01$ & $0.11\pm 0.01$ \\
EW (eV) & $129$ & $107$ & $99$ \\
\cutinhead{Relative Normalizations}
$A_{\mathrm{XIS1}}$ & $1.01\pm 0.01$ & $0.95\pm 0.01$ & $1.01\pm 0.01$
\\
$A_{\mathrm{PIN}}$ & $1.09^{+0.02}_{-0.03}$ & $1.03\pm 0.01$ &
$1.06^{+0.03}_{-0.02}$ \\
\cutinhead{$0.6$--$50$~keV Flux Ratios}
Ionized Reflector/Power-Law & 0.17 & 0.09 & 0.11 \\
Neutral Reflector/Power-Law & 0.05 & 0.03 & 0.04 \\
\enddata
\tablecomments{The Doppler broadening was modeled with the `gsmooth'
  convolution model in XSPEC with the Gaussian width $\sigma(E) =
  \sigma_{\mathrm{6\ keV}}(E/\mathrm{6\ keV})$.}
\end{deluxetable}

As seen in Figure~\ref{fig:fit3}, and in agreement with the expectations
of \citet{hnk04}, the soft excess of \lmc\ is almost
entirely explained by a highly ionized reflection spectrum, and is produced by the
thermal bremsstrahlung emitted by the hot, ionized surface of the
irradiated region. In addition, the broadened
\fe\ line required by \citet{hung10} can be described by a redshifted
and Doppler broadened ionized \fe\ line from the same
reflector. The width of the ionized \fe\ line ($\sim 0.2$~keV) is less
than the one determined by \citet{hung10} ($\sim 0.4$~keV) because the
line is naturally broadened due to Comptonization (see
Figure~\ref{fig:model}). The low-ionization reflector self-consistently explains the
narrow \oeight\ and \fe\ lines \citep{hung10}. However, the reflection
components are a small fraction of the total observed flux. The
low-$\xi$ reflector contributes only $\approx 4$\% of the total flux
for each observation, indicating that this emission region subtends a
consistently small solid angle as seen from the pulsar, and therefore
originates from the outer disk, the accretion stream, or the companion
star. In contrast, the highly ionized reflector contributes
$\approx 10$\% of the broadband flux, but this increases to $17$\%
for Observation 1. As these observations were taken at different
super-orbital periods, as the inner accretion disk precess around the
pulsar, then this changing fraction strongly implies that the ionized
reflector arises from the inner regions of the magnetically-truncated
accretion disk, consistent with the origin of the broadened \fe\ line. Moreover, the change in flux ratio
indicates that our view of the inner reflecting region
evolves with the super-orbital period, as expected for a warped
precessing disk \citep{hung10}. 

The fits require an additional, broad emission component between $0.9$
and 1~\kev\ that does not appear in the reflection model (the best fit
without this feature results in a $\Delta \chi^2\approx+200$). This
emission is well modeled by a broad Gaussian component with a
centroid energy of $0.94$~\kev, $\sigma=0.1$~\kev, and EW$\approx
100$~eV. These properties are significantly different than the weak,
narrow line that \citet{hung10} needed at a similar energy and
tentatively identified as arising from highly ionized Ne. In this
case, a narrow emission line is strongly rejected by the fit, and thus
favors an interpretation that the Gaussian component is modeling a
smooth part of the continuum that is missing from the highly ionized
reflection spectrum. The energy and width of this feature are very
similar to a Gaussian bump in the \xmm\ RGS spectra of \lmc\ analyzed by
\citet{neil09}, who modeled the soft excess
using a variety of different continua and persistently observed an
additional Gaussian component. Thus, this feature is unlikely to be an instrumental artifact, but a true
unmodeled component of the continuum (a similar feature was also found
in the spectrum of Her X-1; \citealt{jg02}). The energy range spanned by this emission feature
encompasses many transitions of Fe~\textsc{xvii}--\textsc{xxiii} \citep[e.g.,][]{brown02}, which, for this
value of $\xi$, are expected to be weak. However, there will likely be a steep
ionization gradient on the surface of the inner accretion disk that is
being parameterized here by only a single ionization parameter. In this
scenario, a single ionized reflection spectrum may not be able to
account for all the spectral complexity at $\la 1$~\kev, a region
which is very sensitive to the ionization parameter
\citep[e.g.,][]{gk10}. Whatever the origin of this feature, it only
contributes $9$\% of the soft excess flux (measured between $0.6$ and
$2$~keV), as compared to $56$\% produced by the ionized reflector.
Therefore, this unmodeled emission does not alter the conclusion that
the highly ionized reflector provides a good description of the soft
excess in \lmc.

The redshift and Doppler broadening of the ionized reflecting gas
indicates that it is subject to strong dynamical effects. One possibility
is relativistic smearing from motions close to the neutron star
\citep[e.g.,][]{fab89}. However, tests with the `rdblur' model show
that good fits can only be obtained if the inner radius of the
reflector was $\sim 20$--$40$~$r_g$ from the star. Identifying this
radius with the magnetospheric radius implies a magnetic field strength
of $\sim 10^9$~G \citep{gl78,gl79,fkr02}, significantly smaller than
the $\sim 10^{13}$~G field strength required to explain the stability
of the \lmc\ pulse period \citep{woo96}. Moreover, if the disk
extended down to $\sim 20$--$40$~$r_g$ the
accretion luminosity from the disk would overwhelm the observed
spectrum unless the accretion rate was $<0.07$ of the Eddington rate,
which is problematic given the observed luminosity of $3\times
10^{38}$~erg~s$^{-1}$. Thus, as indicated by its low covering factor, this gas must originate much farther out
from the neutron star, perhaps from the inner edge of the disk as it
interacts with the magnetosphere. With a magnetic field strength of
$\sim 10^{13}$~G, the magnetospheric radius of \lmc\ is $\sim 2\times
10^8$~cm which would have an orbital velocity of $\sim
10^4$~km~s$^{-1}$ \citep{hung10}, in good agreement with both the
measured redshift and the velocity width at $6.7$~\kev. This conclusion is supported by the
results of \citet{neil09} who uncovered evidence in high
resolution gratings spectra for multiple regions of high-speed ionized gas
flows. It is clear that ionized reflection spectra not only explain
the soft-excess of X-ray pulsars, but allows a probe of the complicated
dynamics associated with the interaction of the magnetosphere and
accretion disk.

\section{Conclusions}
\label{sect:concl}
This Letter presents new reflection models developed for accreting
X-ray pulsars. Thermal bremsstrahlung subject to Comptonization from
an ionized X-ray heated disk is found to be a natural explanation for
the soft excess observed in the spectra of \lmc, and, presumably,
other X-ray pulsars which exhibit \fe\ lines and soft
excesses. Fitting the \suzaku\ data of \lmc\ nicely illustrates the
potential of reflection modeling for elucidating the physics of these
systems. We found that the observed broad \fe\ line could be explained
as originating from a redshifted ($\sim 10^4$~km~s$^{-1}$) highly
ionized reflector subject to Doppler broadening. These velocities are
consistent with those expected from the magnetospheric radius of $\sim
2\times 10^8$~cm \citep{hung10}. Thus, applying this grid of
reflection models to both archival and future X-ray pulsar spectra
will provide important constraints on the ionization state,
metallicity and dynamics at the inner edge of magnetically-truncated
accretion disks.
  
\acknowledgments
The authors thank the anonymous referee for a very helpful report.
This work was supported in part by NSF award AST 1008067 to
DRB. RCH acknowledges \textit{Suzaku} grant NNX08AI17G.

{}

\end{document}